\pdfoutput=1 

\documentclass[12pt]{article}

\usepackage[bookmarks=false]{hyperref}

\usepackage{amsmath}
\usepackage{graphicx}
\usepackage{enumerate}
\usepackage{natbib}
\usepackage{url} 

\newcommand{\blind}{1}

\begin{document}

\def\spacingset#1{\renewcommand{\baselinestretch}%
{#1}\small\normalsize} \spacingset{1}


\if1\blind
{
  \title{\bf Temporal and spatial downscaling for solar radiation}
  \author{Maggie D. Bailey$^{1,2}$\thanks{
    The authors gratefully acknowledge the U.S. Department of Energy Oﬀice of Energy Eﬀiciency and Renewable Energy Solar Energy Technologies Oﬀice, Contract No. DE-AC36-08GO28308}\hspace{.2cm}\\
    Douglas W. Nychka$^1$\\
    Manajit Sengupta$^2$\\
    Jaemo Yang$^2$\\
    Soutir Bandyopadhyay$^1$ \\
    \\
    $^1$Department of Applied Mathematics and Statistics, Colorado School of Mines\\
    $^2$National Renewable Energy Lab\\
}
  \maketitle
} \fi

\if0\blind
{
  \bigskip
  \bigskip
  \bigskip
  \begin{center}
    {\LARGE\bf Building high-resolution gridded solar radiation data under climate change scenarios}
\end{center}
  \medskip
} \fi

\bigskip
\begin{abstract}
Global and regional climate model projections are useful for gauging future patterns of climate variables, including solar radiation, but data from these models is often too coarse to assess local impacts. Within the context of solar radiation, the changing climate may have an effect on photovoltaic (PV) production, especially as the PV industry moves to extend plant lifetimes to 50 years. Predicting PV production while taking into account a changing climate requires data at a resolution that is useful for building PV plants. Although temporal and spatial downscaling of solar radiation data is widely studied, we present a novel method to downscale solar radiation data from daily averages to hourly profiles, while maintaining spatial correlation of parameters characterizing the diurnal profile of solar radiation. The method focuses on the use of a diurnal template which can be shifted and scaled according to the time or year and location and the use of thin plate splines for spatial downscaling. This analysis is applied to data from the National Solar Radiation Database housed at the National Renewable Energy Lab and a case study of the mentioned methods over several sub-regions of continental United States is presented. 
\end{abstract}

\noindent%
{\it Keywords:}  Statistical Downscaling, Global Horizontal Irradiance, Functional Data Analysis, Global and Regional Climate Models
\vfill

\newpage
\spacingset{1.9} 
\section{Introduction}
\label{sec:intro}

Our future climate is uncertain. However, years of observed climate and weather patterns and the development of Earth system models allows for projections of future climate based on different trajectories of greenhouse gas emissions and other atmospheric changes (\cite{flato2014evaluation}). Although the change of global surface temperature has been given the most publicity and attention, it is anticipated that a variety of other atmospheric properties, such as clouds and aerosols, may be affected but the extent is uncertain (\cite{zhang2020effect, mcneill2017atmospheric}). This in turn may change the amount of incoming solar radiation at the surface of the earth. Because these changes are based on an increase in greenhouse gases, which are incremental over long time periods, they are gradual and will evolve over years if not decades. Long term changes in radiation have emerged as an important factor in lifetimes of photovoltaic (PV) facilities. To improve on efficiency and return on investment, new PV facilities are being designed to have lifetimes of 50 or more years and such long-term planning requires consideration of a changing climate. In particular, it is important to site new facilities in areas where the expected total solar radiation will remain consistent, or increase. Additionally, changes within the distribution of solar radiation will affect integration into the electric grid. Typically, planning for PV facility placement has been done based on historical data. However, there is an increased awareness that planning done under historical scenarios is not representative of future climate change scenarios. Therefore, understanding the future of solar resources is vital for this form of renewable energy. 

Solar radiation is among many climate variables simulated by global and regional climate models, the primary numerical tools to understand the earth's potential climatic futures. However, model projections are often available on time and space scales that are too coarse for solar radiation integration modelling, local resource assessment, and other downstream modeling where a finer time and spatial scale is needed (\cite{carvallo2023guide}). As an industrial standard, the National Renewable Energy Lab (NREL) produces the National Solar Radiation Database (NSRDB), giving data on hourly and half-hourly time scales at the 4km level (\cite{sengupta2018national}). This historical data product is used for the type of PV facility placement mentioned above. High-temporal-resolution solar resource data (hourly or sub-hourly resolution) is required for solar PV power generation models to estimate system performance (\cite{Buster2021}). Usually, the resulting solar power generation estimates are used as inputs for capacity expansion and production cost models for solar integration studies and at least hourly-resolution solar climate data is needed to simulate power system operations for future climate scenarios. In contrast, solar radiation simulations from regional climate models are typically available at 25 or 50 km resolution on a daily average time scale.  Therefore, it is of interest to generate solar radiation projections for future years at an hourly time scale and a finer spatial scale to aid in planning PV facilities that may operate during projected changes. In the geosciences, this process of inferring processes at finer scales from coarser information is referred to as \textit{downscaling}. Downscaling projections of climate model data will help characterize how climate variability might affect future solar resource.

Based on this motivation, this work addresses the statistical problem of temporally and spatially downscaling solar radiation, moving from daily average solar radiation values to hourly sequences and from a 20 km to an 8 km grid. At a conceptual level, there are two types of downscaling: dynamical and statistical. Dynamical downscaling is the process of forcing regional climate models (RCMs) using general circulation models (GCMs), which operate at a larger, global scale. Here, the coarse resolution GCM data is used as input boundary conditions for the RCM models. The idea is to derive local climate and weather patterns from large scale climate patterns generated by GCMs. Dynamical downscaling methods have the benefit of being based on physical relationships between variables and can be fine-tuned to regional climate (\cite{mcginnis2021building}). However, there are several drawbacks including being computationally expensive (\cite{solman1999local}), being available only at specific resolutions, and often having systematic biases. Alternatively, statistical downscaling is based on establishing a statistical relationship between large and small scale variables. Statistical downscaling has several advantages including using less computation resources and focusing on a specific variable or use (\cite{benestad2007empirical}). There is also an emerging area of research to downscale physical processes directly from GCMs, though existing methods in this area are sparse (\cite{harris2022multi, buster2024high}).

\subsection{Review of current methods}

Methods for temporal downscaling of solar radiation are often referred to as \textit{synthetic time series generation}. Similar to the classification defined in \cite{Grantham2018}, synthetic time series generation can be generalized into two main categories: (1) interpolating or downscaling from observed solar radiation at equally spaced time intervals or (2) generating time series that have not been observed but are similar in the statistical properties of observed time series. As an example for category (2), consider generating a hourly solar radiation profile for say June 6, 2005 that was not observed but matches the \textit{distribution} of observed hourly radiation at that same day in the same location. 

There are many methods that downscale data as defined by the first definition. Some examples include \cite{fernandez2016} which downscales hourly measurements of solar radiation to time series at a minute scale by matching the closest day between the hourly input of observed data and a generated hourly time series. 
There are many other studies that temporally downscale according to this first definition (\cite{Buster2021, frimane2019nonparametric, zhang2018stochastic, Grantham2017, skartveit1992probability}).


The second approach is less widely studied but important for future projections. \cite{Graham1990} developed a simple algorithm that disaggregates the hourly clear sky index, $k_t$, into an additive model: $k_t = k_{tm} + \alpha$. Here, the clear sky index is a measure of atmospheric transparency for incoming solar radiation. They identified sets $\{k_{tm}\}$, a trend component, and $\{\alpha\}$, a noise component, for all values of $k_t$, the daily average clear sky index, and assessed the variability observed in $\alpha$ by conditioning according to $k_t$. In this way, the method is similar to an analog method, which uses observed data at the desired spatial or time scale as the downscaled data (\cite{zorita1999analog}), but only for the errors, $\alpha$. \cite{aguiar1992tag} developed the Time-Dependent, Autoregressive, Gaussian (TAG) Model for hourly time series generation based on the daily clearness index as input. The TAG algorithm claims better consistency compared to \cite{Graham1990}, which may be due to more locations that were used in its development. \cite{grantham2016nonparametric} proposes an additive model to generate hourly synthetic times series of GHI data. The model includes a Fourier series component, autoregressive component, and a stochastic component. The stochastic component bins errors according to solar elevation and zenith angle. Then, synthetic series are created by drawing from these bins to add stochastic error for the given solar angle and elevation at the hour being modeled. This method is similar to an analog method which essentially uses observed data as the final simulation. Finally, the distribution of the synthetic solar radiation is adjusted to match observed data so that statistical measures are realistic. \cite{Grantham2018} proposes a method to  generate synthetic daily and hourly sequences by first developing a model for the deterministic component, using Fourier series, and then adds a stochastic component, where the white noise term is sampled from a Beta distribution, which builds off a nonparametric bootstrapping method introduced in \cite{boland2008time}. \cite{mora1998multiplicative} propose a method that downscales monthly average clearness index to an hourly time scale using multiplicative autoregressive, moving average models. 

The studies mentioned above all follow a process of first defining a trend component and then adding noise. This is the general approach taken in this paper, however our method differs by leveraging the functional form of hourly solar radiation time series and emphasizes adding spatially correlated daily patterns for sets of nearby locations. Coherent patterns across space are a more realistic representation of variability from large scale cloudiness in a region.



\subsection{Novelty and paper structure}

We propose a functional data approach to capture the diurnal pattern of hourly solar radiation and also account for the variation that is not predictable from a daily summary. Although solar radiation is an unlimited resource, its  main drawback is intermittency, primarily due to clouds. Clouds may affect the stability and power quality of a PV plant so it is important to capture some amount of intermittency as it affects the final power output of a PV system (\cite{jhr2010impacts}). Therefore the goal is to downscale daily solar radiation reflecting this variability but also maintain observed spatial and temporal correlations. Our model is similar to previous methods in that its structure includes a trend and noise component:
\begin{equation}
    y(h,d,s) = GHI(d,s) \times T(h,d,s) + \sum_{j = 1}^J \phi_j(h)u_j(d,s)  + \varepsilon(h,d,s)
    \label{intro_mod_fda}
\end{equation}

\noindent where here we model $y$, the solar radiation for hour $h$ of day $d$ at location $s$. The first term, $GHI(d,s) \times T(h, d, s)$, is the trend component. $T(h,d,s)$ is a diurnal pattern or template adjusted by $GHI(d,s)$, the daily solar radiation average for day $d$ at location $s$. Thus, $T$ has the constraint that $\sum_{h = 1}^{24} T(h,d,s) = 1$. The second term is the noise component, where we establish a set of basis functions, $\phi_j(h)$, for $j = 1, \ldots, J$, weighted by $u_j(d,s)$, spatially coherent coefficients predicted for location $s$ and day $d$.

Our method preserves daily total solar radiation to ensure projections of solar radiation neither inflate nor decrease daily totals. Additionally, it is capable of producing multiple realizations that reflect the expected variability at finer space and time scales. Although previous work also quantifies this variability it does not address the coherence of solar radiation as a space-time process where variation is similar at close locations and close times. The method described in this paper is adaptable to any location, day of year, and daily average solar radiation value where appropriate observed data is available. Finally, the method proposed in this paper aims to be explainable to non-statisticians and the broader community, efficiently scalable to the entire contiguous United States (CONUS), and implementable within a high-performance computing environment in addition to statistically accurate for simulating hourly GHI. Some steps may be statistically justifiable with more complicated methods. However, to achieve a balance of interpretability, accuracy, and efficiency, we opt for a simpler approach at several steps and these situations are noted throughout the paper.

The rest of the paper is organized as follows: the data used to develop the model is described in Section~\ref{sec:data}; in Section~\ref{sec:methods} we show how functional principal component analysis (FPCA) motivates the use of a template model; the diurnal template model is presented in Section~\ref{sec:diurnal_model}; Section~\ref{sec:spatial_downscaling} describes our approach to spatial downscaling; in Section~\ref{sec:scaling_conus} we introduce a method to scale the model to work across CONUS; validation and results are provided in Section~\ref{sec:results}; and, lastly, Section~\ref{sec:conclusion} provides concluding remarks and outlines future work.

\section{Data}
\label{sec:data}

\subsection{National Solar Radiation Database}

\begin{figure}
    \centering
    \includegraphics[width=.45\textwidth, trim={0 0 0 0},clip]{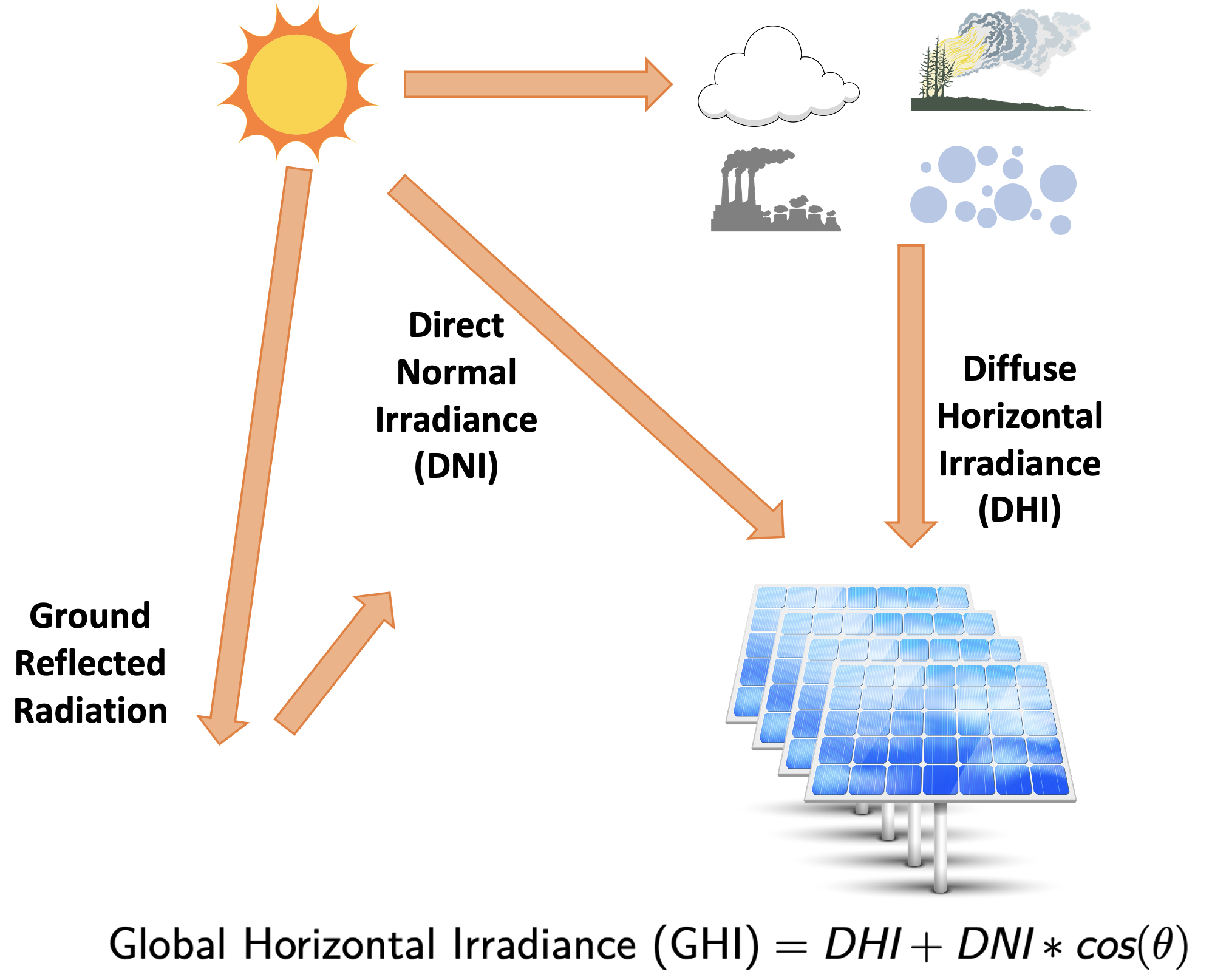}
    \includegraphics[width=0.5\textwidth]{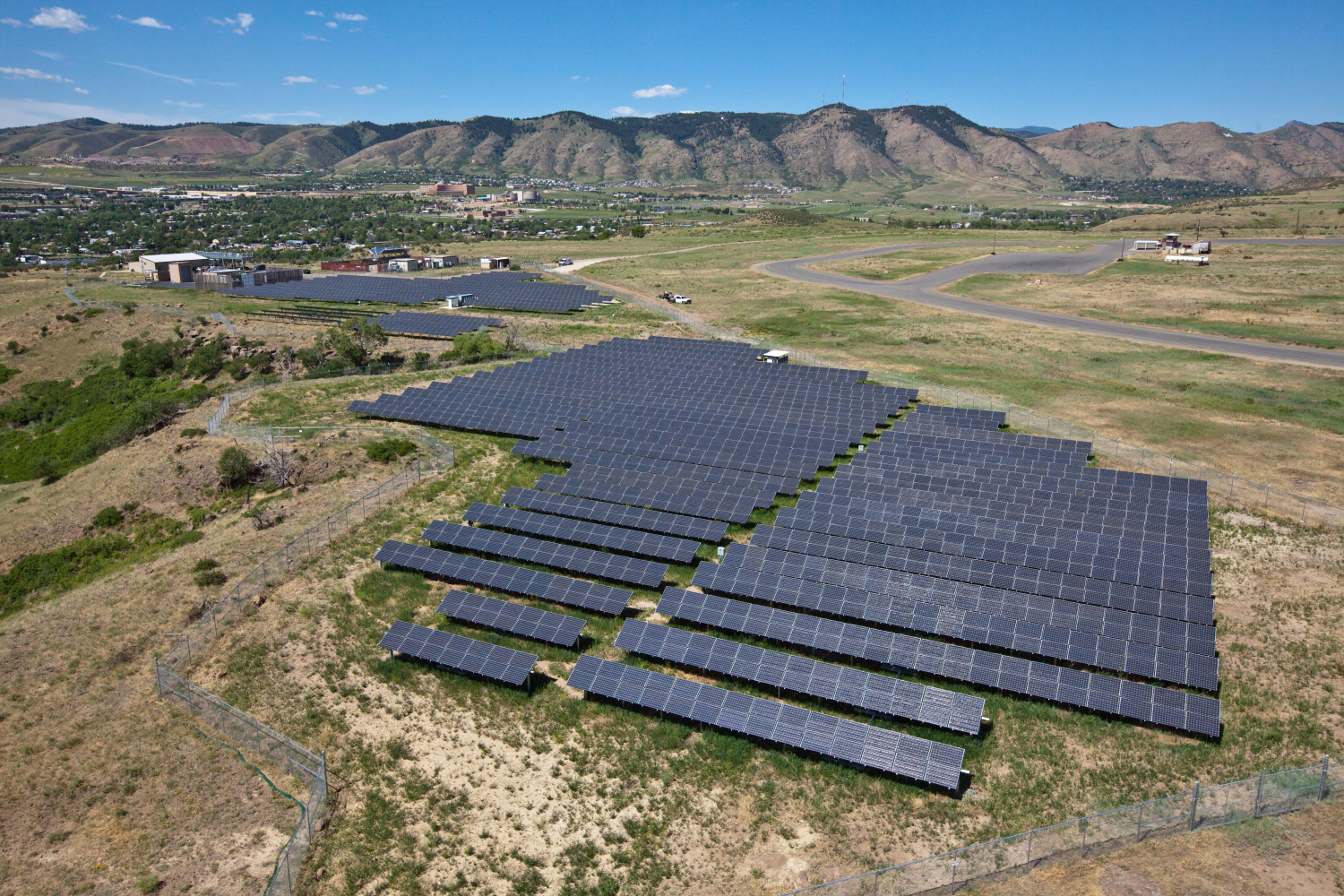}
    \caption{Left: DNI is the amount of radiation received by an area perpendicular to the sun's rays while DHI may be scattered by clouds, pollutants, wildfire smoke, and aerosols, among other atmospheric components. Here, $\theta$ is the solar zenith angle. Right: a photovoltaic array near NREL on South Table Mountain in Golden, Colorado (\cite{doe2010array}).}
    \label{fig:solar_radiation_explained}
\end{figure}

The NSRDB includes a solar radiation data product complete in time and space for CONUS over the period 1998-2022 at the 4 km grid resolution. (See \cite{sengupta2018national} and the references therein for more details). The NSRDB is widely used by a variety of agencies, including local and federal governments, utility companies, and university researchers. Although primarily used for energy related applications, its uses have been extended to other areas such as the connection between solar exposure and health effects (\cite{zhou2019compilation}). It contains hourly and half-hourly measurements for the three most common solar radiation variables: global horizontal irradiance (GHI), direct normal irradiance (DNI), and diffuse horizontal radiation (DHI), in units of $Wm^{-2}$. Additionally, it provides data for a subset of years for most countries across the globe. Figure~\ref{fig:solar_radiation_explained} provides an illustration to how DNI and DHI contribute to GHI. Solar radiation data is calculated using the Physical Solar Model that takes as input cloud properties derived from satellite measurements to calculate DNI and DHI (\cite{sengupta2018national}). GHI is then derived from these two variables. The NSRDB product for GHI and DNI has been validated and shown to be accurate to within 5\% (GHI) and 10\% (DNI) when compared to surface observations (\cite{habte2017evaluation, yang2021validation}).  For the purposes of temporal downscaling, however, the data set is averaged to a grid resolution of 20 km to be comparable with regional climate model output at a similar resolution. Section~\ref{sec:spatial_downscaling} will discuss spatial downscaling from this grid to an 8 km grid.

\section{Functional Principal Component Analysis}
\label{sec:methods}

\begin{figure}[t]
    \centering
    \includegraphics[width=0.8\textwidth]{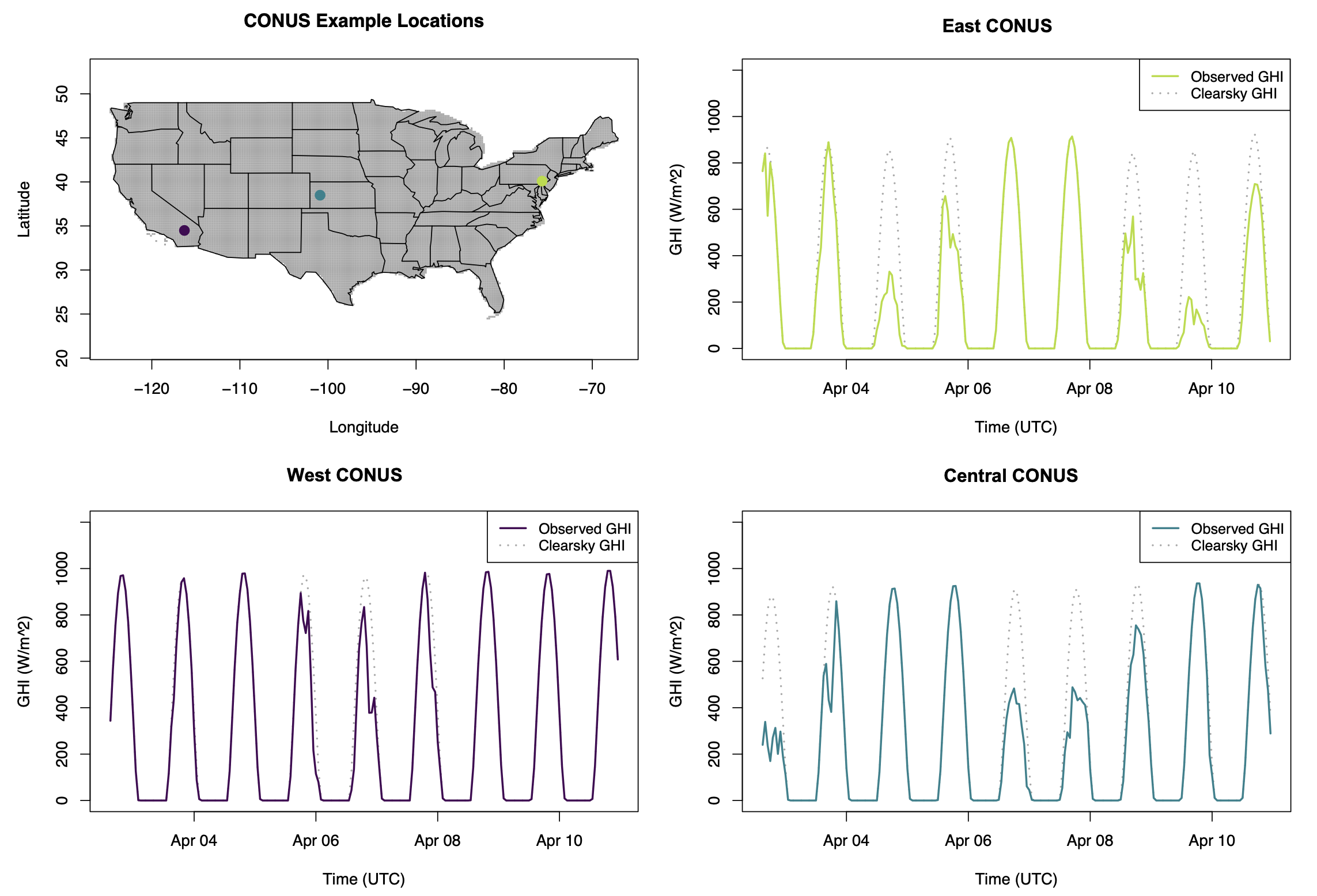}
    \caption{Example time series for three locations across the US in the month of October. The dotted grey line shows the clearsky hourly profile for each day.}
    \label{fig:conus_ts_example}
\end{figure}

One can justify a functional data approach by considering functional principal components of a set of daily solar radiation curves. Figure~\ref{fig:conus_ts_example} is a set of eight time series of solar radiation for three locations in CONUS. Within each set of time series, we see clearsky days, fully cloudy days, and intermittently cloudy days with periods of both cloudy and clearsky conditions. Clearly these time series show a strong diurnal pattern similar to a sinusoid or a parabola. On cloudy days, the clearsky curve is reduced and adds a stochastic element to the hourly time series. Similar to classical principal component analysis, functional principal component analysis (FPCA) identifies the main modes of variation departing from a mean curve within a sample of functional curves (\cite{shang2014survey}). 




Let $X$ be a $k \times h$ data matrix of hourly solar radiation profiles, where $k = N \times T$,  across $N$ locations, $T$ total days and $h=24$ hours. After centering $X$, resulting in $X^c$, the matrix can be decomposed using the SVD and then further written:
\begin{equation}
    X = \mu^c +  \sum_{j=1}^h \delta_j \boldsymbol{u}_j\boldsymbol{v}_j
    \label{eq:svd_x_expansion}
\end{equation}

\noindent where $\mu^c$ is the mean profile, or vector of column means of $X^c$, and $\delta_1 \geq \delta_2  \geq \ldots \geq \delta_j \geq \ldots \geq \delta_h$. The right-singular vectors $\boldsymbol{v}_j$ represent the major modes of variation. For our purposes, the vectors $\boldsymbol{v}_j$ will be identified as the basis functions in (\ref{intro_mod_fda}). Each $\boldsymbol{v}_j$ is weighted by $\delta_j$ so that $\delta_1\boldsymbol{v_1}$ can be thought of as the function showing the largest mode of variation from a mean profile, $\delta_2\boldsymbol{v_2}$ the second major mode of variation, and so on. In this sense, $\delta_1\boldsymbol{v_{1,h}} = \phi_1(h)$ as seen in (\ref{intro_mod_fda}). The variance explained by the first $J$ basis functions can be assessed through analyzing $\delta_j^2$. Often, not all basis functions are needed to recreate a profile that accurately captures a majority of the variability from the observed profile.


\begin{figure}[t]
    \centering
    \includegraphics[width=.7\textwidth]{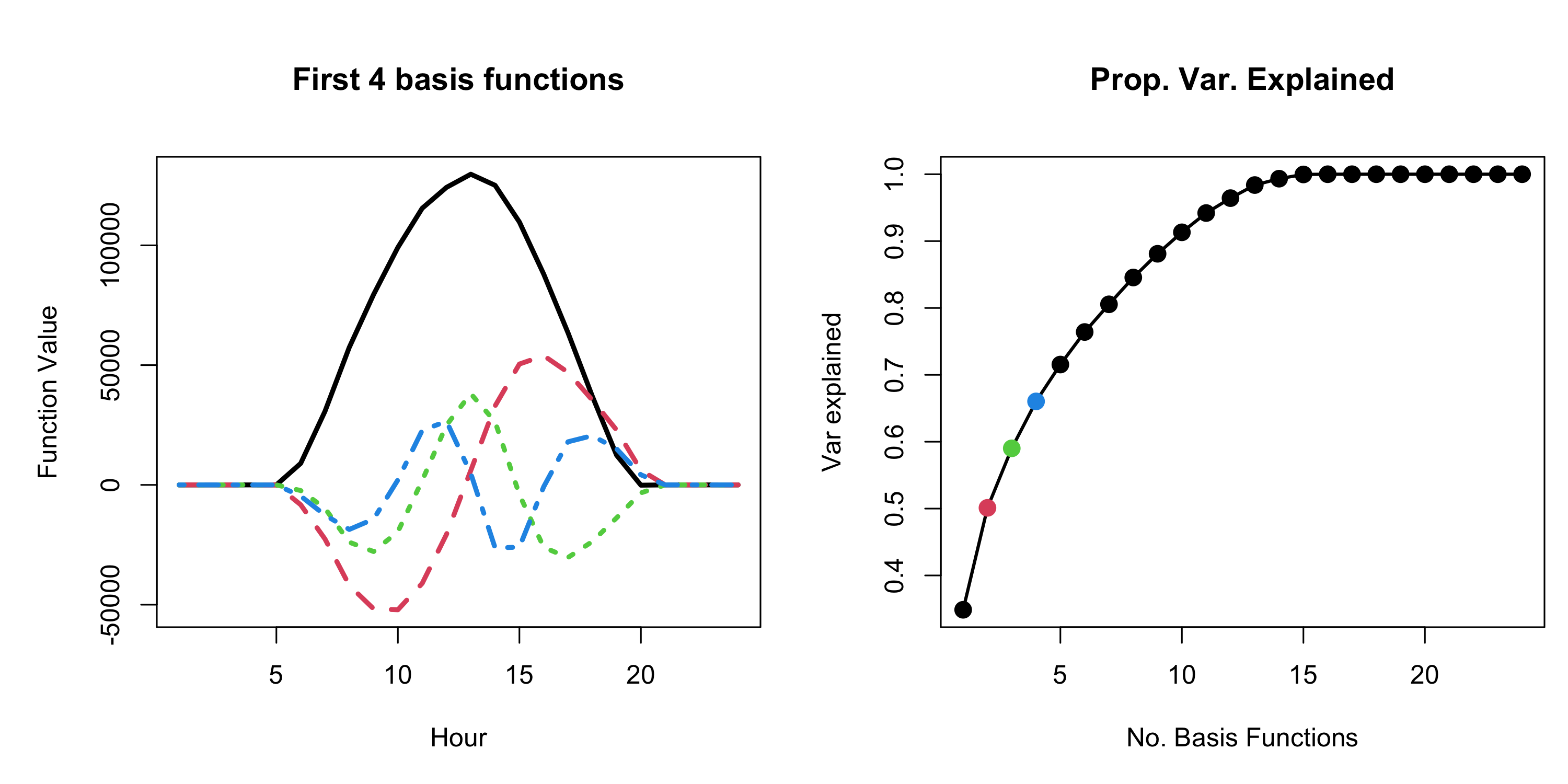}
    \includegraphics[width=.4\textwidth]{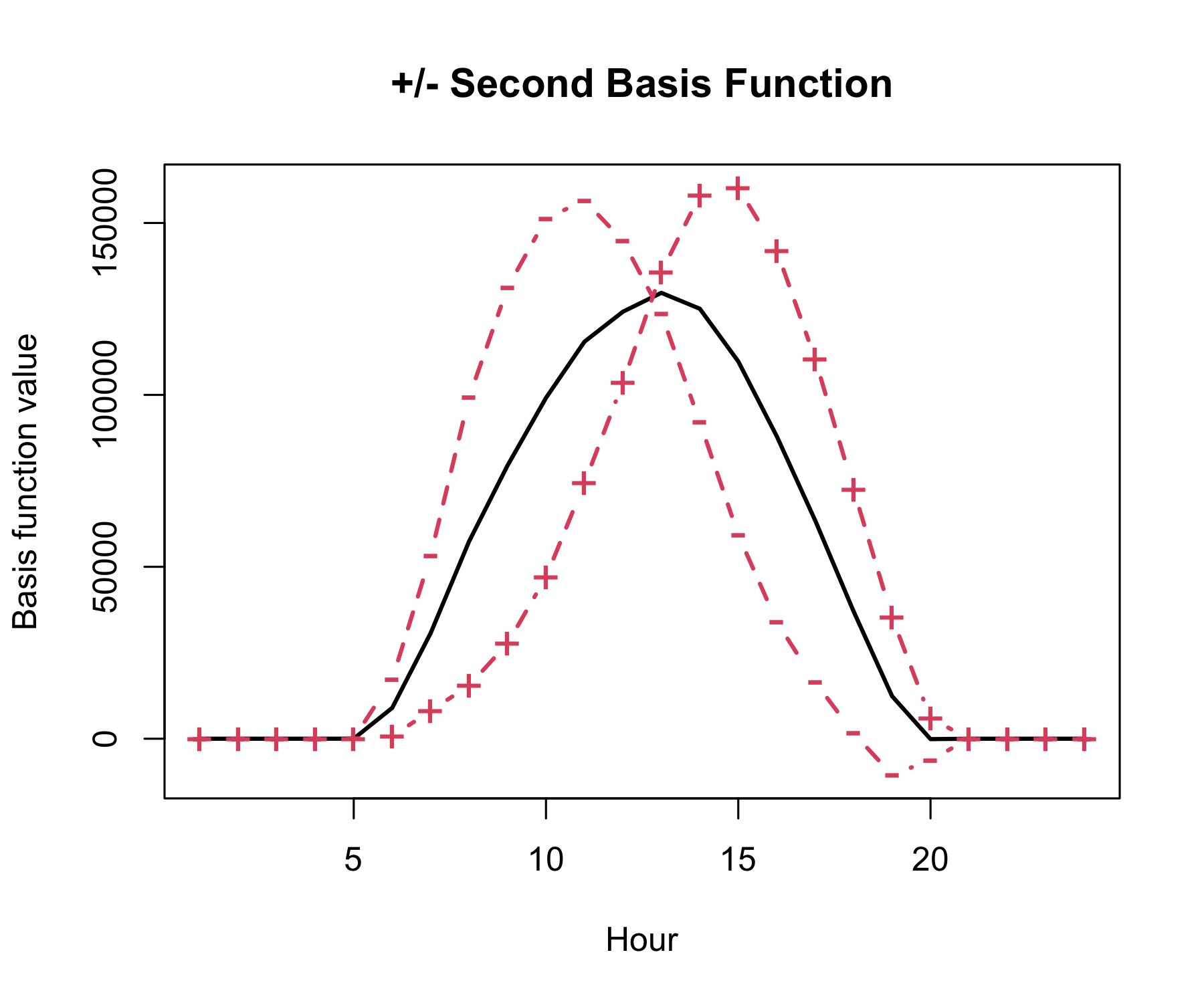}
    \includegraphics[width=.4\textwidth]{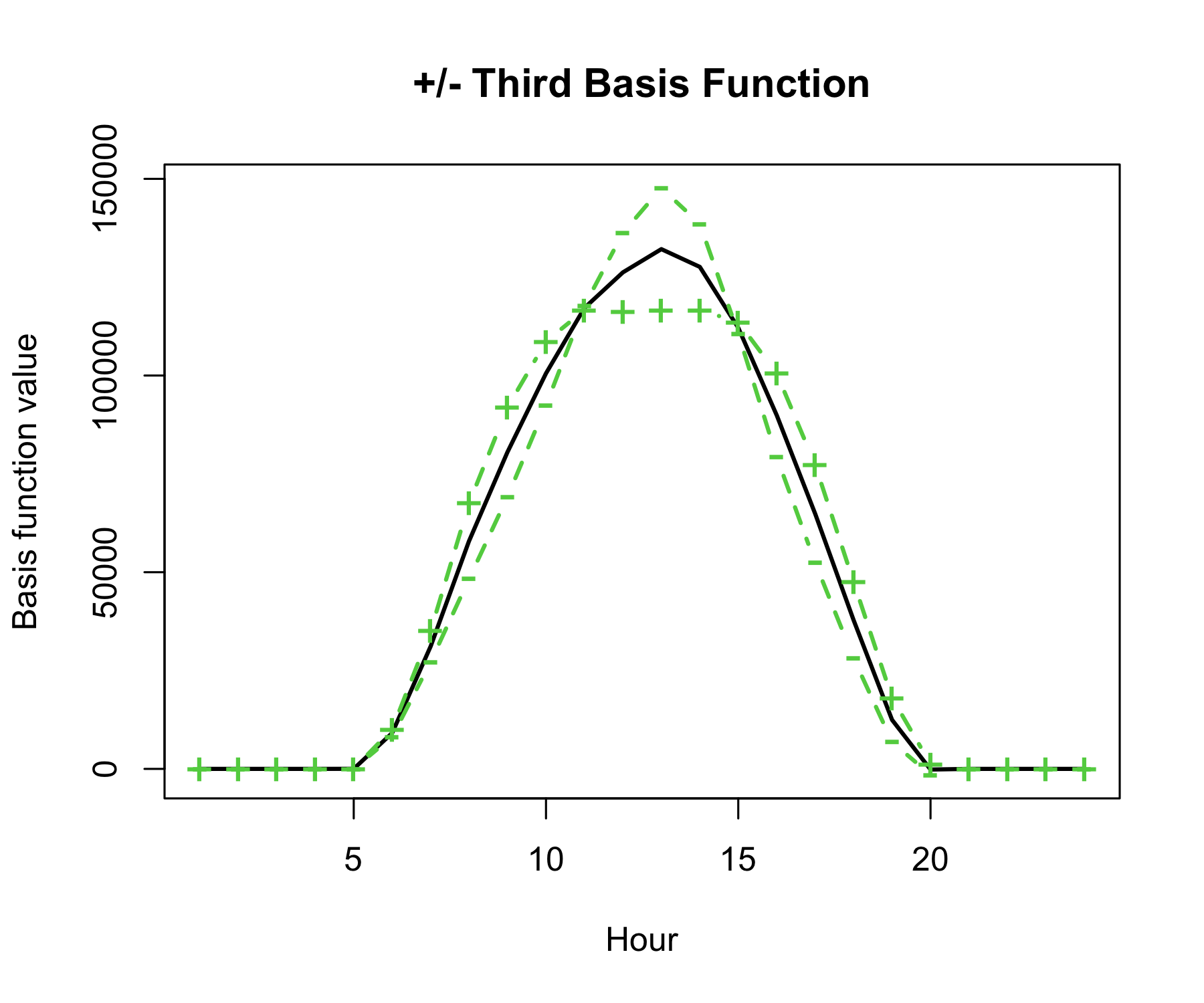}
    \caption{Results from FPCA on a centered data matrix for GHI in June. Top left: first four basis functions for June. Top right: proportion variation explained by basis functions, colors match the lines in plot on the top left. Bottom row: first basis function plus/minus the second and third, indicating how the second and third basis functions affect variability in the functional curve.}
    \label{fig:eda_fpca_xcent}
\end{figure}

An example of the resulting basis functions ($\boldsymbol{v}_j$) for the month of June can be seen in Figure~\ref{fig:eda_fpca_xcent}. Data for the month of June comes from the NSRDB across 23 years of data (1998-2020) and 1,070 locations in the sub region containing California. The top left plot shows the first four basis functions for the month of June and the top right plot shows the total variability explained in the first $J$ basis functions. The first basis function explains close to 40\% of the variability in the observed profiles and that 5 basis functions explain more than 70\% of the variability. The bottom row shows \textit{plus-minus} plots for the second and third basis functions. Plus-minus plots are an exploratory visualization tool showing how the individual basis functions combine to produce different diurnal patterns. Here, the second basis function controls the time at which peak solar radiation occurs including controlling for the time of local noon. The third basis function controls a dip or peak in solar radiation during mid-day. More generally, the third basis function either adds clouds in the middle of the day or in both the evening and morning. An example of this would be measuring solar radiation in a valley, with mountains blocking early morning and late afternoon sun. 


In this example the first basis function explains more than 40\% of the diurnal variability. Moreover, the next two basis functions can be interpreted as modifying this baseline shape in interpretable ways. This suggests that the first basis function can be used as a template, or function, for all days and locations which can capture a large portion of the variation seen in the diurnal patterns through shifting and scaling. The template approach ensures positive values, a key feature of solar radiation, and allows for the constraint of the daily average. The additional basis functions can be considered noise additions that shift the first basis function in way that may simulate the variability due to clouds. The addition of the second, or more, basis functions will help explain hourly deviations from a clearsky day. Here we focus on using four basis functions as this explains close to 70\% of the variability of the diurnal profile. In order to appropriately weight the additional basis functions, we consider the spatial coherence of the basis function coefficients and propose a spatial model for this feature. The ideas explored in this section inform the model proposed in the next section, which introduces the template model and method for adding hourly variability.

\section{Diurnal Template Model}
\label{sec:diurnal_model}

This section introduces a functional data analysis approach to downscale daily average GHI data to hourly averages. Instead of additive basis functions that are described in the previous section, a fixed template is shifted and scaled to capture the hourly radiation for different locations and seasons. As mentioned in the introduction, the model aims to match daily total GHI that will later be provided by regional climate models.

\subsection{Building a Universal Template}
\label{subsec:building_univ_template}


We start with a function that is representative of radiation based on clear sky conditions. Denote this function as $g^c(h)$ which is normalized so that $\int g^c(h)dh = 1$. Operationally $g^c$ is estimated from the hourly data and interpolated using a cubic spline. However, $g^c$ based on a physical model would also be appropriate. 


\subsection{Adjusted Template}

The template $g^c$ is calculated from observed clear sky GHI profiles across the domain and normalized, acting as a universal template. Therefore it needs to be shifted to match observed clear sky GHI solar radiation profiles for a given location and day of year. Given the clear sky template $g^c$, the mean function for any location and season, we propose the following model to shift the template to predict mean hourly GHI for the $i^{th}$ location:
\begin{equation}
    GHI_{i} = GHI(s_i,d) g^c( \tau_i(h - c_h) - \beta_i + c_h) + \varepsilon_i\label{eq:template_model}
\end{equation}

\noindent where $GHI(s_i,d)$ for $i = 1, \dots, N$ controls the height of the profile and is the daily total GHI given by the observations or future RCM. This value is provided directly by climate model data and the NSRDB. The parameter $\beta_i$ shifts the template left or right, or earlier or later, along the x-axis and controls the time of peak GHI (solar noon) and is in the units of hours. For example if $\beta_i = 0.1$ then the profile is shifted $60\times 0.1 = 6$ minutes forward in the day.  $\tau_i$ is a scale parameter and controls the width, or length of day, of the template for the $i^{th}$ location. Both $\beta_i$ and $\tau_i$ are estimated using nonlinear least squares by month. Finally, the parameter $c_h$ is the average time of solar noon for a set of locations and season, calculated using observed data, and is important for scaling to CONUS, described later. 



\begin{figure}
    \centering
    \includegraphics[width=0.45\textwidth]{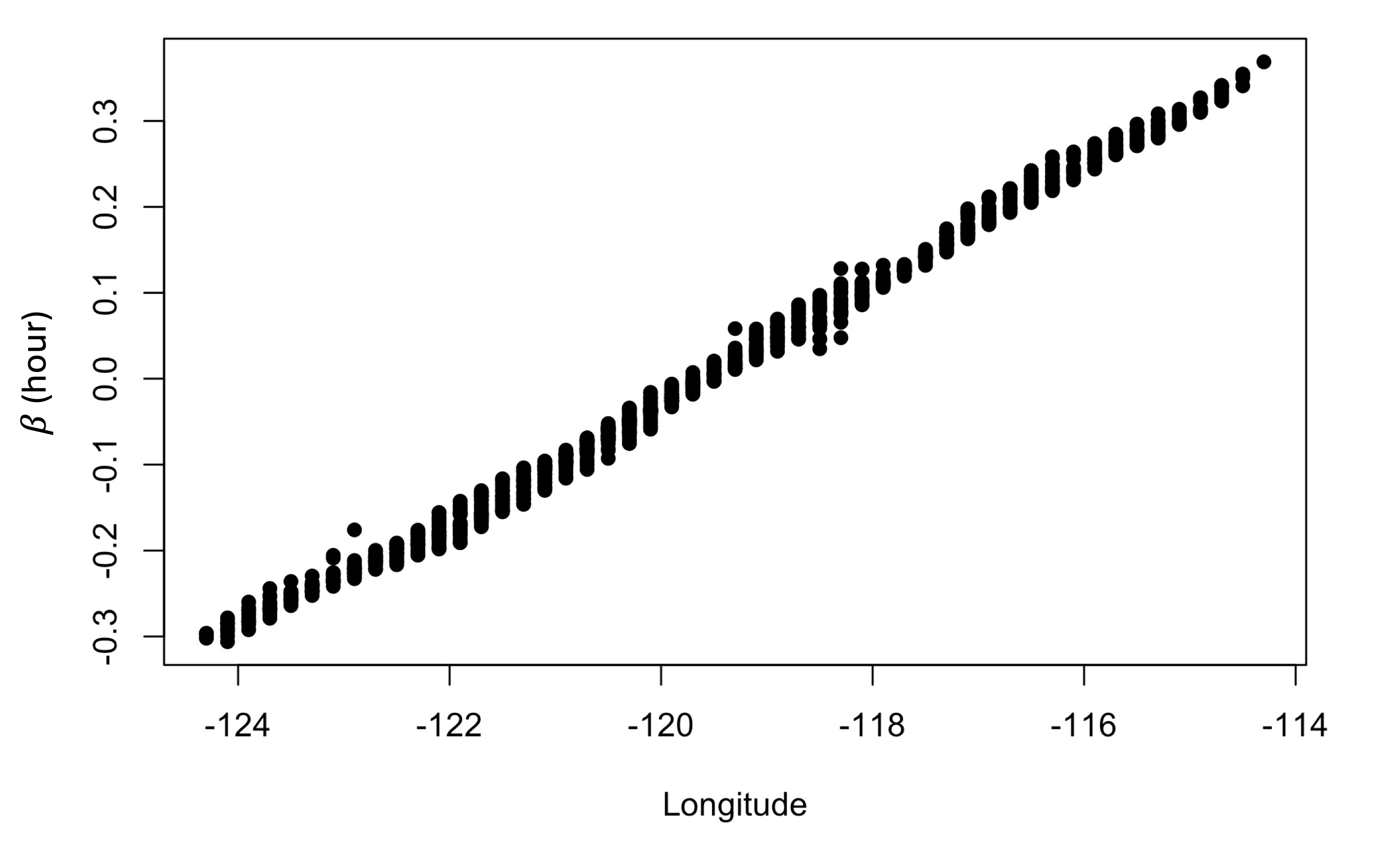}
    \includegraphics[width = 0.45\textwidth]{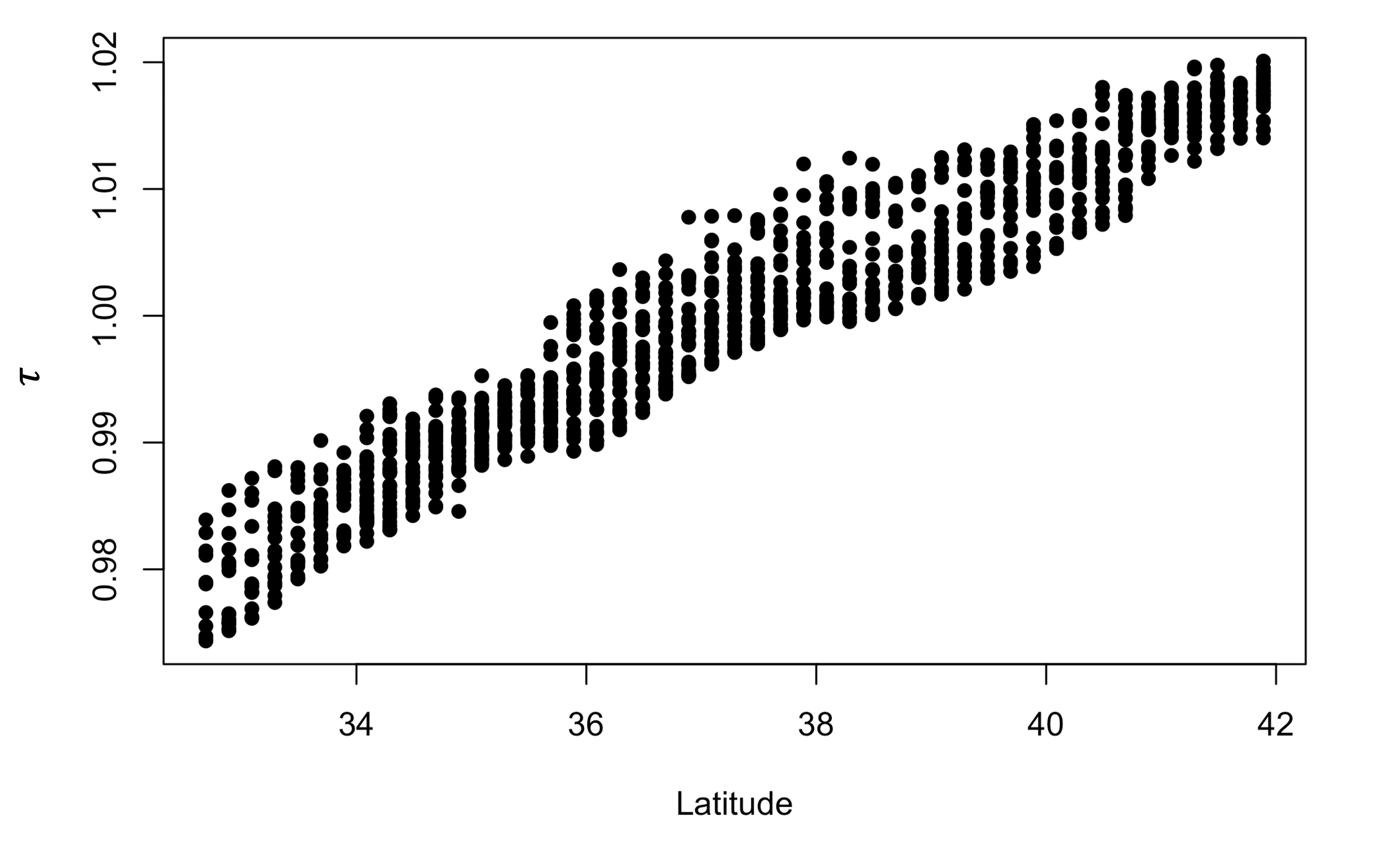}
    \caption{Estimates of $\beta$ and $\tau$ plotted against longitude and latitude, respectively, for the month of August over a subset of locations in western CONUS. $\beta$ is in units of hours and $\tau$ scales the length of daylight. }
    \label{fig:beta_tau_geography}
\end{figure}

The parameters $\beta$ and $\tau$ depend on geography and season and they are interpretable given a location and month. The time of solar noon is dependent on longitude and so $\beta$ can be related to longitude. Length of day is a function of latitude and so $\tau$ can be estimated based on latitude. Because these relationships shift during the year, $g^c$ is adjusted for each month of the year across the designated locations. Resulting estimates for $\beta_i$ and $\tau_i$ are shown in Figure~\ref{fig:beta_tau_geography} for the month of August and a subset of locations over western CONUS. It is clear that there is a strong linear relationship between longitude ($z_{1,i}$), latitude ($z_{2,i}$) and the two parameters, respectively, and we can leverage this linearity to predict $\beta_i$ and $\tau_i$ through the linear models:
\begin{align*}
    \beta_i &= \gamma^\beta_0 + \gamma^\beta_1z_{1,i} + \varepsilon^\beta \\
    \tau_i &= \gamma^\tau_0 + \gamma^\tau_1z_{2,i} + \varepsilon^\tau.
    \label{eq:tau_lms}
\end{align*}

We now adjust the height of the template through the parameter $\alpha$. As mentioned previously, the universal template has been normalized to the sum to 1 and was done to ensure the the final adjusted template will equal the daily total GHI given by the NSRDB. 



\subsection{Adding spatially correlated clouds}
\label{subsec:noise_addition}


Although the mean model based on a template is effective in representing the mean profiles there is substantial variability about this function. This variability can be imagined through the differences seen between the clear sky profiles and observed hourly data as in Figure~\ref{fig:conus_ts_example}. Although the user can gain basic information from the mean profile, additional variability that is important for hourly variation is clearly missing.

To fit the noise portion of the model, we first identify the residuals from (\ref{eq:template_model}), which is fit by month. The residual matrix, $E$, created in the same way as described in (\ref{eq:svd_x_expansion}), is decomposed into its singular value decomposition
\begin{equation}
    E = UDV^T 
    \label{eq:resid_svd}
\end{equation}

\noindent resulting in a the matrix $U$ containing basis function coefficients over space and time. Note that we are subsetting data by sets of locations and by month and thus $E$ contains residual data for a set of locations and one month of the year.  The resulting residual basis functions resemble the second, third, and fourth basis functions shown in Figure~\ref{fig:eda_fpca_xcent}.

Referencing the downscaling model at Eq.~\ref{intro_mod_fda}, the problem is then how to predict or simulate the coefficients by day and location. Solar radiation is a spatio-temporal process so we expect daily average GHI to be similar at nearby locations and less similar over larger distances. Therefore, it is reasonable to expect the coefficients to exhibit spatial coherence. 

\begin{figure}[t]
    \centering
    \includegraphics[width=.85\textwidth]{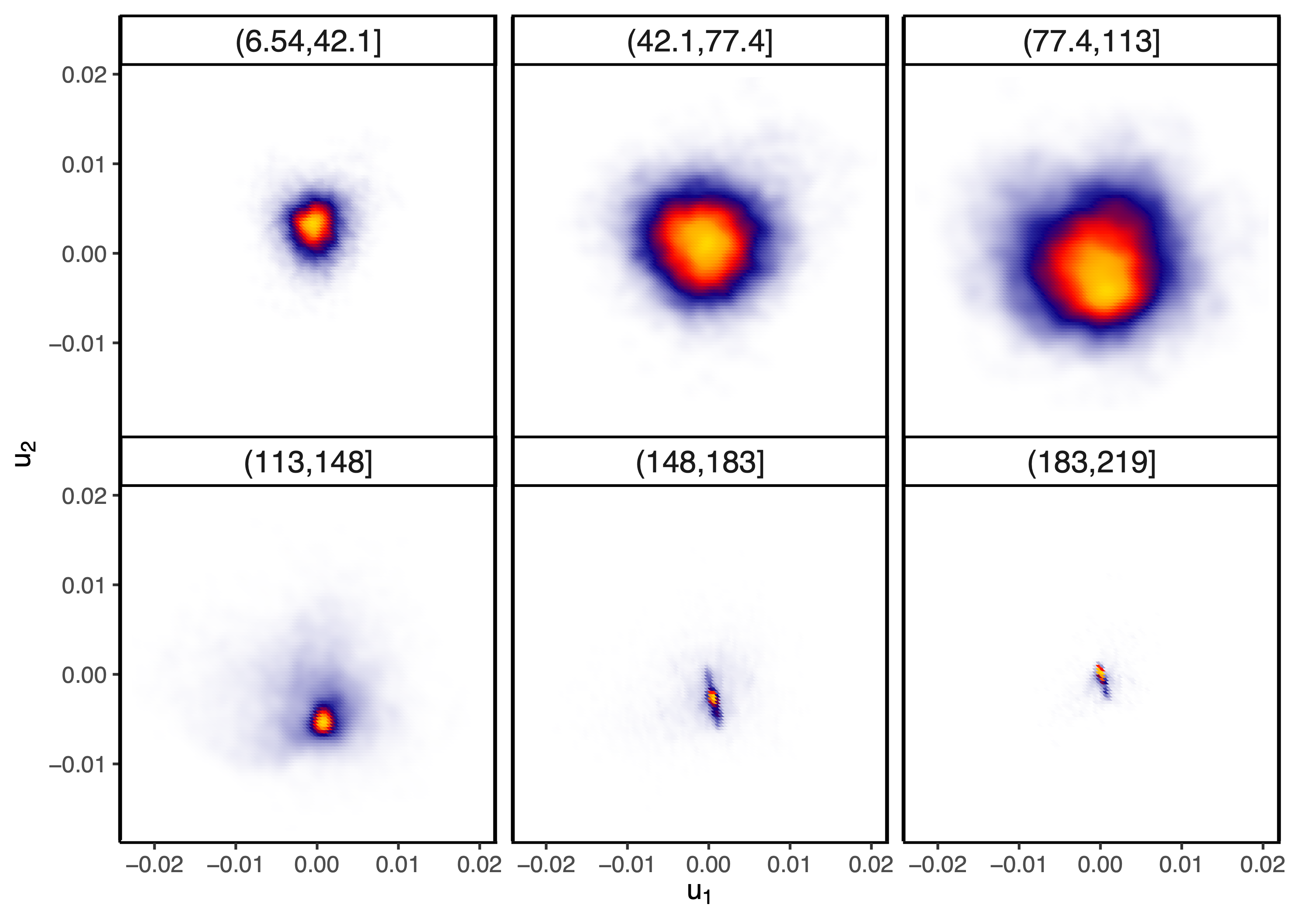}
    \caption{The first two residual basis function weights for the month of January plotted against each other for six bins of GHI, starting at a low daily average GHI on the top left and ending with high daily average GHI on bottom right. The color indicates the density of points, estimated using a two-dimensional, bi-variate normal kernel density function.}
    \label{fig:u1_vs_u2}
\end{figure}

\begin{figure}[t]
    \centering
    \includegraphics[width=0.9\textwidth]{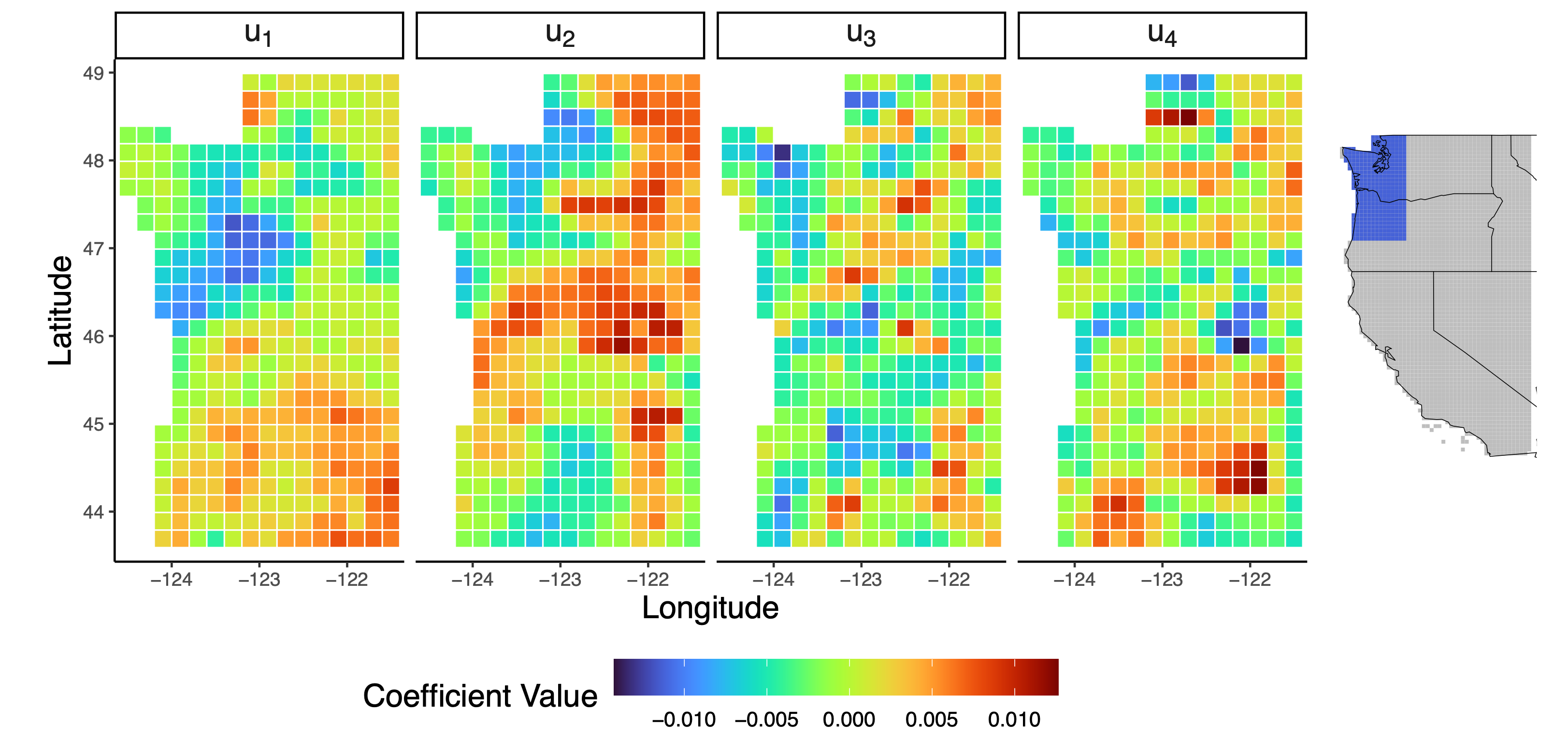}
    \caption{First four basis function weights for a set of locations in Nortthwest CONUS for a day in October.}
    \label{fig:u1_thru_u4}
\end{figure}

\subsection{Basis Function Coefficient Properties}
\label{sec:bf_coeff_preoperties}

Exploratory analysis on the basis function coefficients show dependence conditioned on daily average GHI values. Figure~\ref{fig:u1_vs_u2} are density plots of paired coefficients conditioning on daily average GHI within a particular range. Clearly, depending on the bin, the variability in the distribution of the coefficient changes. The final bin (bottom right plot) shows a scatter plot between the first and second coefficient for GHI $\in [183, 219]$. Most of the coefficients are concentrated around zero. These values of GHI represent days with little to no variability due to clouds and so most of the coefficients are close to zero. As the daily average GHI value decreases, the variability of each conditional distribution tends to increase, indicating a greater contribution to the diurnal cycle as it deviates from the clearsky mean. A second result from this plot is that empirically the coefficients appear to be independent of one another. This relationship is also apparent among other pairs of coefficients ($u_1(d,s), \ldots, u_4(d,s)$). Spatial correlation of the diurnal coefficients are exemplified in Figure~\ref{fig:u1_thru_u4} in Northwest CONUS for a day in October. There is clear spatial correlation up through $u_4$ suggesting simulating the coefficients over space for a given location and day may be appropriate for simulating cloud patterns.

To model the diurnal coefficients in space, we propose a model that assumes different variances based on GHI and independence among coefficients but dependence between different spatial locations. We estimate the conditional variance for the distribution $[\boldsymbol{u}_{j}|GHI(\boldsymbol{s},d)]\sim N(0, \sigma_j^2(GHI(\boldsymbol{s},d)))$, where $\boldsymbol{u}_{j}$ are the coefficients for $\phi_j(h)$ over locations $\boldsymbol{s}$ and the daily average solar radiation value, $GHI(\boldsymbol{s},d)$. We further define 
\begin{equation*}
    \boldsymbol{u}^*_j = \frac{\boldsymbol{u}_{j}}{\sigma_{j}^2(GHI(\boldsymbol{s},d))}.
\end{equation*}



 We now consider a spatial model for $\boldsymbol{u}^*_{j}$ of the form:
\begin{equation}
[\boldsymbol{u}^*_{j}|GHI(\boldsymbol{s},d)] = x(\boldsymbol{s},d)\beta + f_j(\boldsymbol{s},d) + \varepsilon_j(\boldsymbol{s},d)
\label{eq:u_coeff_spatial_model}
\end{equation}

\noindent where $x(\textbf{s}, d)$ contains a field of daily average GHI values in $\boldsymbol{s}$ for day $d$, $f_j(\boldsymbol{s},d)$ is a mean zero Gaussian process with covariance function $k(\boldsymbol{s},\boldsymbol{s})$ and $\varepsilon_j(\boldsymbol{s},d)\sim N(0,\sigma^*_j)$. Simulations of $\boldsymbol{u}^*_{j}$ resulting from Eq.~\ref{eq:u_coeff_spatial_model} are re-scaled through multiplication of  $\sigma_{j}^2(GHI(\boldsymbol{s},d))$.



The simulated coefficients can now be used to simulate hourly GHI values as
\begin{equation}
    GHI(\boldsymbol{s},h,d) \sim GHI(\boldsymbol{s},d) \hat{g}^c(\cdot) + \sum_{j = 1}^4 \boldsymbol{u}_{j}(\boldsymbol{s},d)\phi_j(h). \label{eq:final_ghi_hat_mod}
\end{equation}

\noindent The values in $GHI(\boldsymbol{s},h,d)$ are a field of hourly simulated GHI data, downscaled from daily averages, with spatially correlated noise components. $GHI(\boldsymbol{s},d)$ is a vector daily average GHI values over $\boldsymbol{s}$ given by a climate model. Simulated data may exceed observed GHI data values, so we impose a final processing step by constraining simulated values to be within physically plausible, observed ranges of hourly GHI day given by the NSRDB.

\section{Spatial Downscaling}
\label{sec:spatial_downscaling}

Solar radiation data on the 20 km grid is often considered too coarse for subsequent applications in PV system planning. This work proposes spatially downscaling to the 8 km grid from the 20 km grid using thin plate splines (TPS) (\cite{hutchinson1983new}). This work as a published data product should meet several goals: it must be flexible for gridded, nonstationary spatial data, adaptable for a variety of geographic areas and sizes, easily implementable at scale for CONUS, and understandable for atmospheric scientists and community users who would be using these methods and data downstream. Spatially downscaling is done using publicly available software in the R package \texttt{fields} (\cite{nychka2015package}). There are other methods that may be considered state-of-the art that can handle larger datasets for spatial downscaling, such as multi-resolution basis functions or Bayesian methods (\cite{paciorek2006spatial, katzfuss2013bayesian, nychka2019package}).  However we adopt a simpler TPS method. This method fits the function $h(\cdot)$ through minimizing the residual sum of squares with the inclusion of a roughness penalty:
\begin{equation*}
    min\sum_{i=1}^N (y_i - h(\boldsymbol{x_i}))^2 + \lambda \int\int \left[\left(\frac{\delta^2h}{\delta x_1^2}\right)^2 + 2\left(\frac{\delta^2 h}{\delta x_1 \delta x_2}\right)^2 + \left( \frac{\delta^2 h}{\delta x_2^2} \right)^2\right]\delta x_1 \delta x_2
\end{equation*}

\noindent where $\lambda$ controls the penalty between soothing and wiggliness of $h$, measured by a penalty function seen in the second term.  In this work, $\lambda$ is selected through maximum likelihood. To justify using TPS, we first downscale the 20 km hourly NSRDB data to the 8 km  grid using TPS. We then compare root mean squared error between the true 8 km hourly data and the predicted 8 km hourly data to the hourly standard deviation of the predicted data. Justification of this approach and results are given in Section~\ref{sec:results}.

\section{Scaling to CONUS}
\label{sec:scaling_conus}

The goal of this project is a temporal downscaling model for all of CONUS. The model described in (\ref{eq:final_ghi_hat_mod}) suggests a single universal template for a given domain. This is impractical given the heterogeneity in terrain and climate across CONUS. We therefore propose a computationally efficient method for implementing the proposed model for CONUS while addressing any nonstationarity of the parameters $\beta$ and $\tau$ as well as covariance parameters estimated for (\ref{eq:u_coeff_spatial_model}) across time and space. 

We divide CONUS into a number of rectangular tiles, where the adjusted template and final downscaled data will be predicted. Each tile is nested in a ``super tile" that adds a border and overlaps with adjacent tiles. The tile centers define locations for \textit{estimating} the universal template, linear models for $\beta$ and $\tau$, and the residual noise model using all pixels in the parent super tile. The size of the super tile depends on a user defined margin that is chosen to be a percentage of the width in degrees of the target tile. For this application, we chose the margin to be 0.4, or 40\% the width of the target tile, to ensure an ample number of pixels to enforce some continuity. All model training is done on the super tile while prediction is done on the target tile.  Implementing the super tile approach aims to ensure correlation between neighboring tiles in space while avoiding edge effects. It also allows for parallellization to improve computational efficiency as the model can be estimated independently for each tile. 

A similar approach is taken to reduce boundary effects over time. The template and parameters are re-estimated for each month of the year resulting in 12 templates and parameter models for each super tile. We add a time buffer of 10 days around each month to estimate the diurnal template and parameters for a given month. 


\begin{figure}
    \centering
    \includegraphics[width=.5\textwidth]{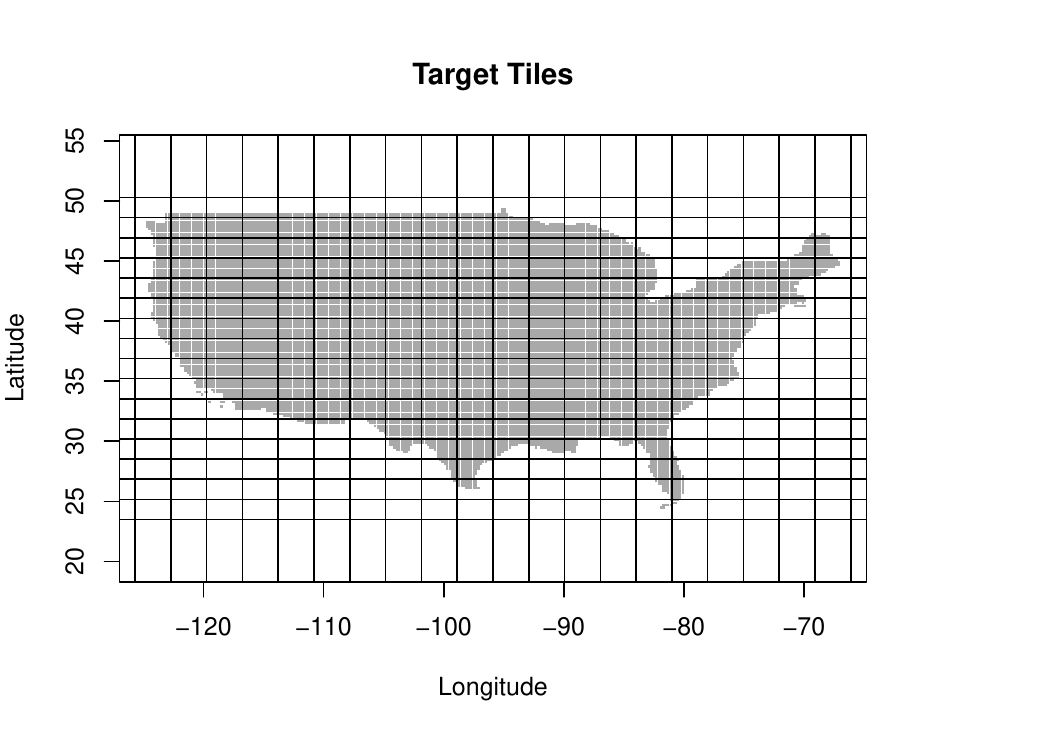}
    \includegraphics[width=.35\textwidth]{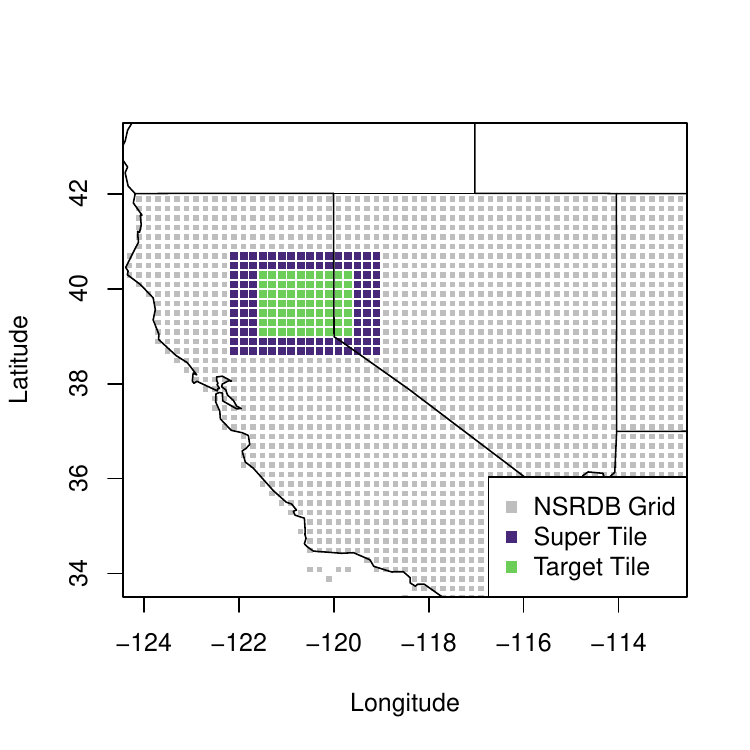}
    \caption{Left: CONUS split into 20$\times$16 = 320 target tiles. Many tiles do not have locations as they are over a body of water. Right: An example of a super tile and target tile over California and western Nevada. The width of the super tile is 140\% the width of the target tile.}
    \label{fig:conus_tile_examples}
\end{figure}


We expect the estimated covariance parameters for the spatial random effect estimated in the noise addition model to vary from tile to tile for a given month. That is, we expect nonstationarity of the spatial model described in Section~\ref{subsec:noise_addition} for CONUS. Therefore, we fit a Gaussian process for the covariance parameters in each super tile. The overlapping nature of these regions results in smooth covariance parameter estimates across space. The super tile approach should naturally account for this nonstationary and further smoothing will ensure natural transitions across the domain. 


\section{Results and Validation}
\label{sec:results}

\subsection{Temporal Downscaling}

\subsubsection{Analysis Design}

\begin{figure}
    \centering
    \includegraphics[width=.65\textwidth]{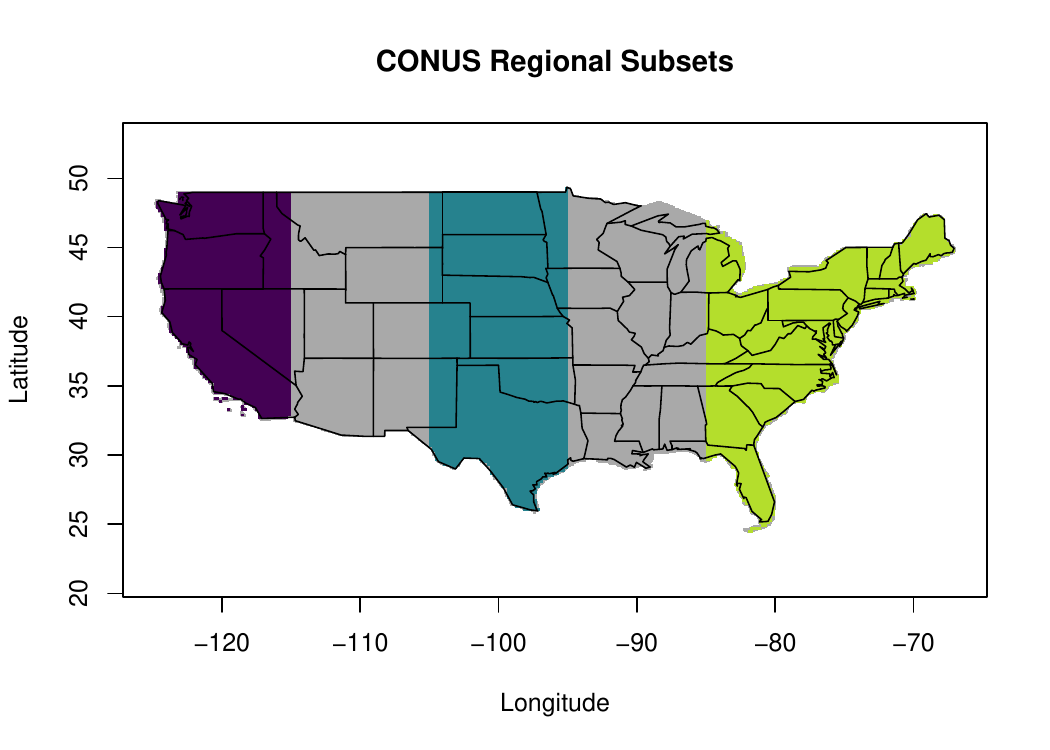}
    \caption{The three sub-regions of CONUS used for fitting the model for years 2000-2009 in January, April, July, and October. The purple shows the western region, teal the central region, and green the eastern region.}
    \label{fig:conus_subregions_fitting}
\end{figure}

The model described in Section~\ref{sec:diurnal_model} has been fitted for four months, January, April, July, and October, and for the years 2000-2009 to illustrate seasonal differences in the fit throughout the year. This results in approximately $10\times 30 = 300$ days for fitting. Additionally, the model is fitted in three separate study regions of CONUS - western, central and eastern region - each with over 3,000 pixels (Figure~\ref{fig:conus_subregions_fitting}). Each sub-region is split into 150 tiles using the scaling setup described in Section~\ref{sec:scaling_conus}. This number was chosen as it resulted in the best balance of computational efficiency for the region size and smoothness of resulting data across space.

Hourly simulations are generated for the same time frames and locations, then compared to observations at the same resolution. It is important to note that it is not expected that the simulated data will match observed data, so calculating a metric such as the root mean squared error between the final simulated data and the observed data will have little meaning in this analysis. However we expect similar distributional properties of the downscaled simulated data to those of the NSRDB, such as intra-daily variability and other summary statistics that help describe the distribution of GHI. We therefore consider several metrics to assess how well the simulated data capture similar patterns and variability as the observed data. 

We compare (1) the distribution of clearsky index by hour between simulated and observed, (2) the density of GHI values by hour of the day to assess if similar values of GHI are represented in each hour of the day by region, and (3) the hourly time derivative of the simulated and observed GHI for daylight hours. 

The time derivative provides a measurement of intra-daily hourly variability and a metric for how well the simulated data captures intermittent sunlight due to atmospheric conditions such as clouds, which is an important measure for estimating PV production. There are other standard measurements of variability available, such as the variability index (\cite{stein2012variability}). This measurement compares hourly variability against clearsky GHI, acting as an error measurement against the clearsky GHI, however we found the time derivative to be a more intuitive measurement of hourly variability. The daily total GHI for each region and month is also assessed to ensure that this method does not artificially inflate or underestimate the total amount of incoming daily solar radiation. 

We validate the spatial auto-correlation of the downscaled data by comparing semivariograms of NSRDB data and downscaled data. Semivariograms are calculated for daylight hours for the four months mentioned above and three regions seen in Figure~\ref{fig:conus_subregions_fitting} and for each daylight hour. For each distance or bin, the 0.25, 0.5, and 0.75 quantile of the semivariogram values are compared between the observed and simulated data. 

The remainder of this section reports results for the sub-region of western CONUS for the four months considered. Other regions of CONUS showed similar results for the same months (see supplemental information).



\begin{figure}[t]
    \centering
    \includegraphics[width=\textwidth, trim={0 5cm 0 0},clip]{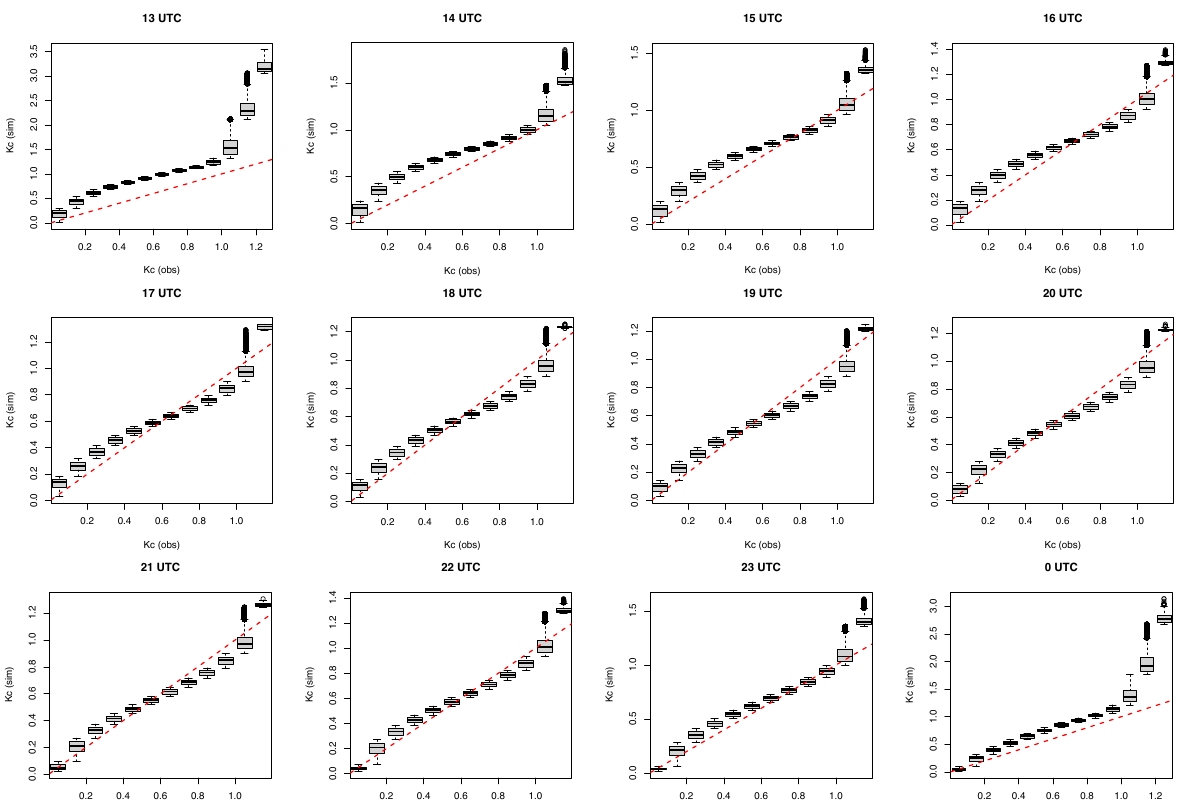}
    \caption{Hourly comparison of $k_c$ values for the western CONUS sub-region in April. The red dashed line indicates the 1-1 line. }
    \label{fig:kc_bplot_west_Jan}
\end{figure}

\subsubsection{Hourly Clearsky Index}

First we consider an hourly comparison of $k_c$ values, the ratio between GHI and the associated clearsky value, for daytime hours in the month of January, plotted in Figure~\ref{fig:kc_bplot_west_Jan}. Early morning and later evening hours show the most disagreement indicating poor agreement in the proportion of $k_c$ values for these hours. Some values are greater than 1 which may be a result of using an empirically derived hourly average clearsky GHI data set to calculate $k_c$. However, for hours towards the middle of the day, simulated $k_c$ closely align with observed values, indicating a strong agreement between the two. In the case of April, shown 
here, the model is slightly underestimating the values for $k_c$ but overall there is a strong agreement. The poor alignment with early and late hour values has two qualifications. Studies have shown that data from the NSRDB is not as reliable for hours when the solar zenith angle is high. The solar zenith angle is measured as the angle between the sun's rays and the vertical direction. Therefore, the data we have for these early and late hours may not be as reliable for fitting and predicting the diurnal template model. Second, most solar power production occurs when the solar zenith angle is low, around local noon time. During the central hours of the day the model is producing good alignment with observed $k_c$ values. 


\begin{figure}
    \centering
    \includegraphics[width=\textwidth]{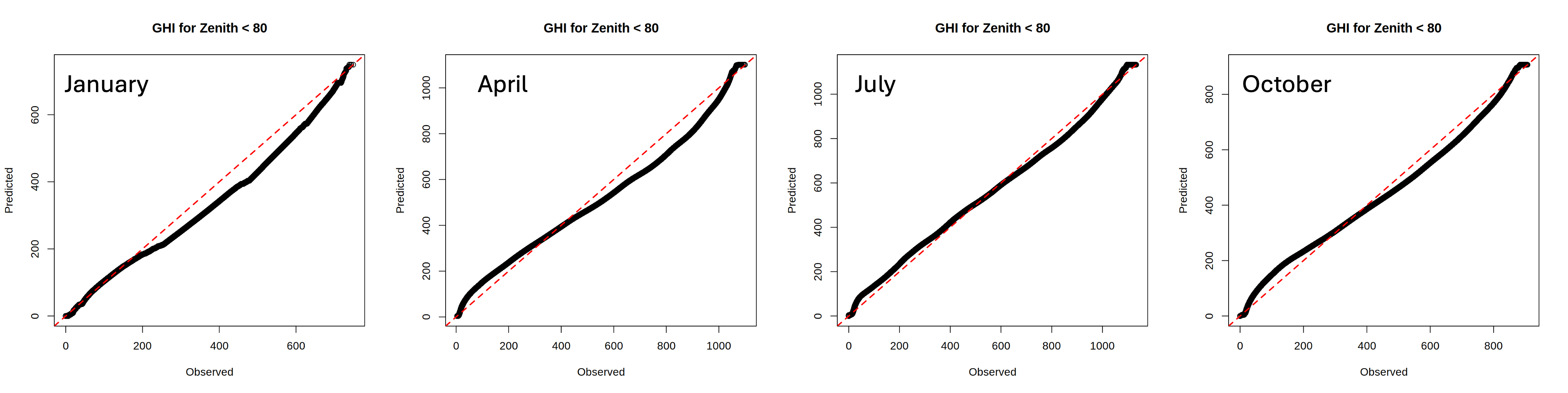}
    \caption{Quantile-quantile plots for hourly GHI values in western CONUS for hour where the sun has a zenith angle less than $80^\circ$, or when the sun is higher in the sky.}
    \label{fig:ghi_zenith_qqplot}
\end{figure}

Considering low solar zenith angle, Figure~\ref{fig:ghi_zenith_qqplot} compares densities of GHI values for hours where the zenith angle is less than $80^\circ$, or when the sun is high in the sky. These quantile-quantile plots indicate strong alignment for values of GHI at these hours when most solar power production occurs showing that similar values for GHI are represented in the downscaled hourly data set.




\subsubsection{Time derivative of GHI}

\begin{figure}
    \centering
    \includegraphics[width=\textwidth]{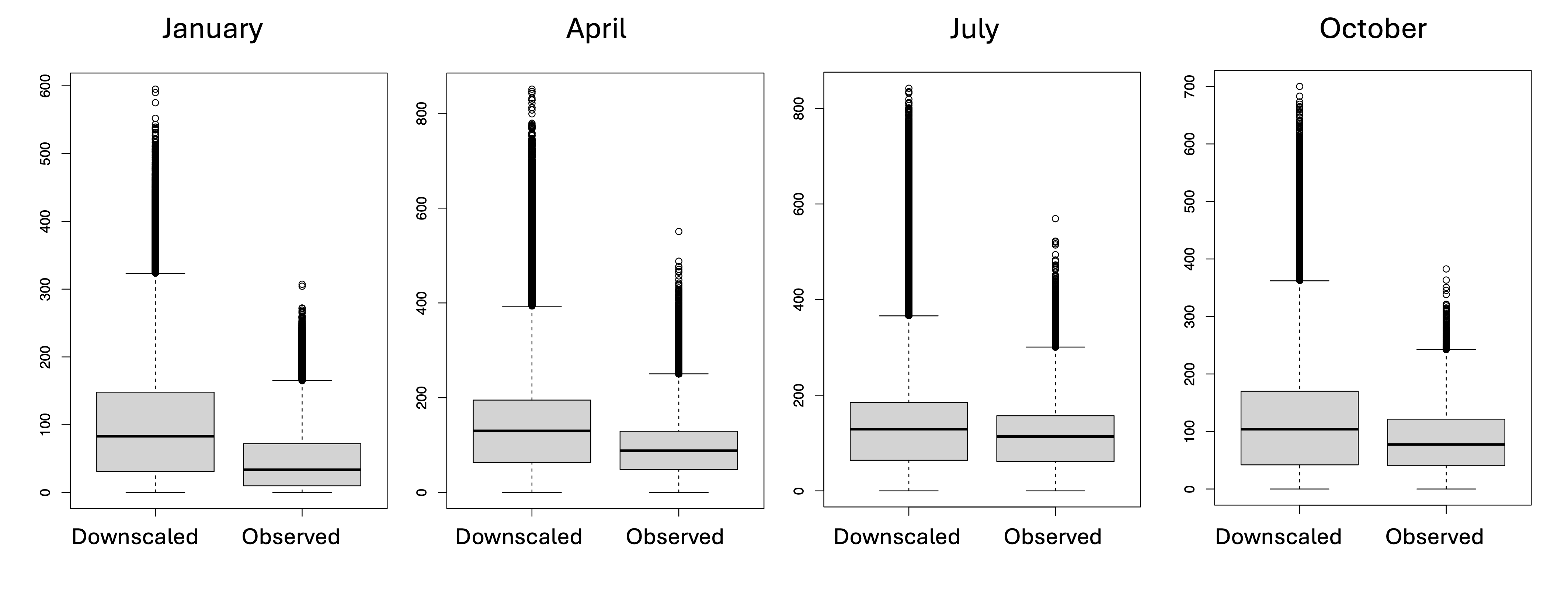}
    \caption{Time derivative comparison for downscaled and observed data for western CONUS in 2009. }
    \label{fig:time_deriv_west_conus}
\end{figure}

Hourly time derivatives for daylight hours are plotted in Figure~\ref{fig:time_deriv_west_conus} for the four months considered in western CONUS. The largest disagreement between the two data types is the overall spread of the data, including outliers, where the simulated shows a wider range of hourly time derivatives than the observed data. July and October have the closest agreement in the range and inter-quartile range where January shows the largest discrepancy. Similar trends were seen across other months and regions of CONUS (see supplement).

It is promising that, overall, the simulated data shows a similar distribution for the time derivatives meaning similar rates of change in GHI are preserved, with the exception of January. This indicates that large increases or decreases in hourly GHI, for example when a cloud might be passing overhead affecting PV production, are generally represented in the simulated data. This also supports using only four residual basis functions to help explain most of the intra-daily variability within the simulated hourly time series.

\subsubsection{Daily Total GHI}

\begin{figure}
    \centering
    \includegraphics[width=\textwidth, height = 1.75in]{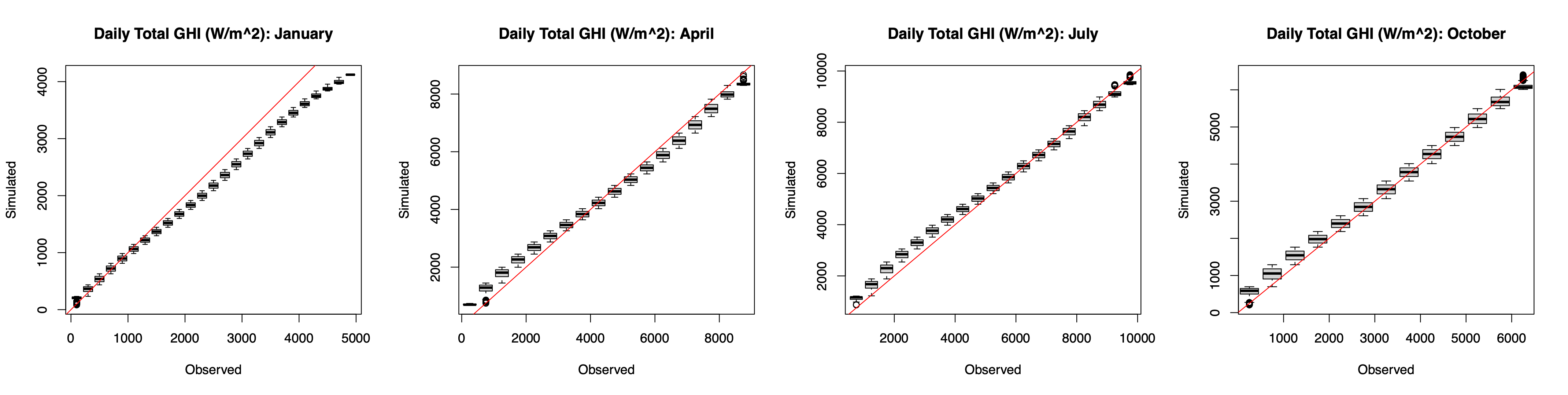}
    \caption{Daily total GHI for observed (x-axis) and simulated (y-axis) GHI data for 2000-2009 for the western CONUS study region. The red line indicates the 1-1 line.}
    \label{fig:daily_tot_ghi}
\end{figure}

Figure~\ref{fig:daily_tot_ghi} shows the daily totals for four months in western CONUS for 2000-2009. We expect the daily totals to closely align as the value $GHI(\boldsymbol{s},d)$ is the given daily total from the observations and the residual basis functions are centered at zero. There is close agreement for all months. However, in the month of January for this region we see that the simulated daily totals slightly under-estimate the observed daily totals for larger daily total values, or for typically clear days. This is due to the model adding variability, not necessarily creating clouds, where the observations may be influenced by cloud patterns. The months of April and July see a mismatch for smaller daily total values while October shows a strong alignment for all values of GHI. Similar results for additional regions for these months can be found in supplemental information.

\subsubsection{Validating spatial dependence}

\begin{figure}[t]
    \centering
    \includegraphics[width=\textwidth]{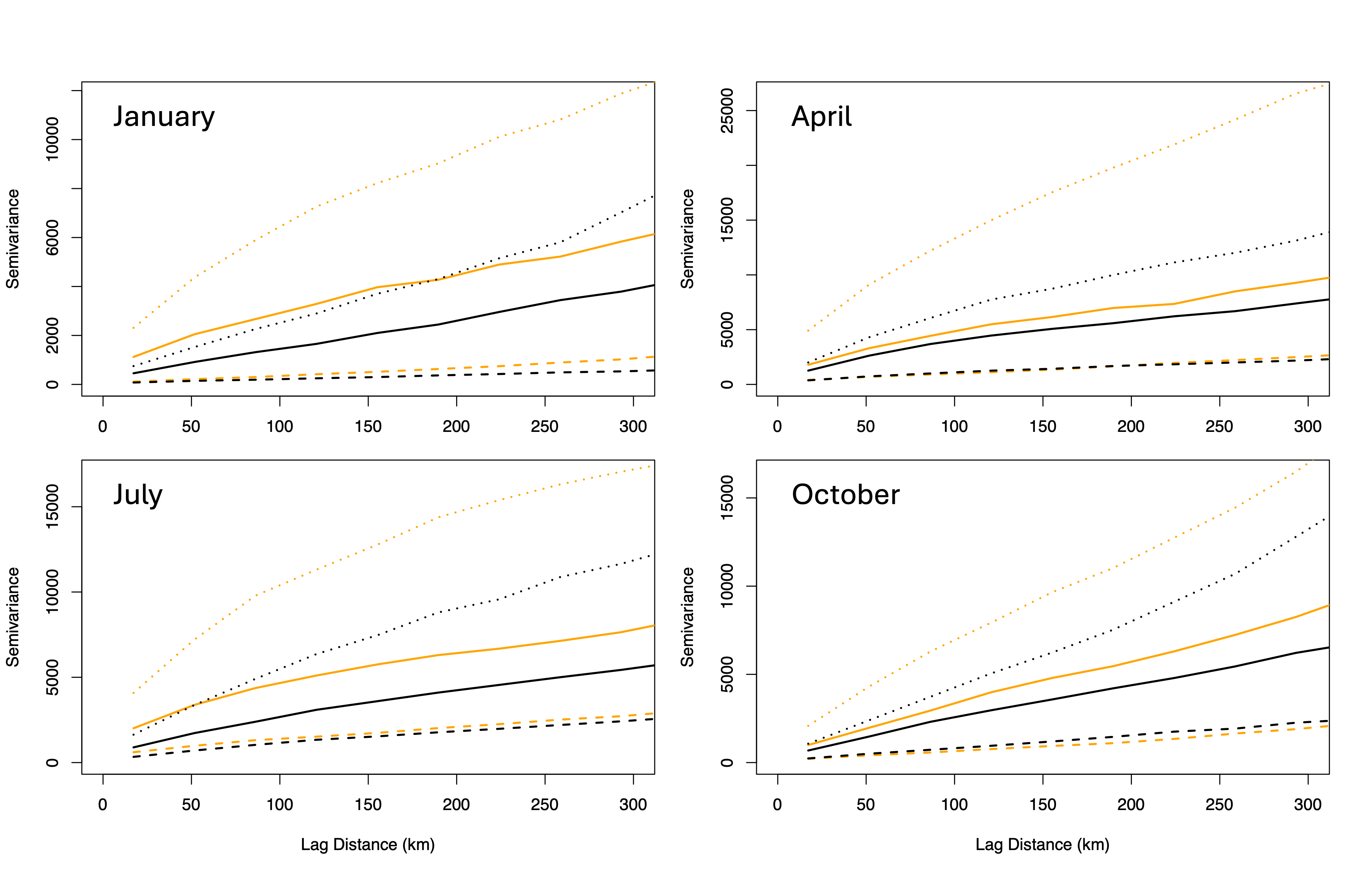}
    \caption{Semivariograms for observed (orange lines) and simulated (black lines) data in western CONUS for 2009. The semivariance is calculated for each daylight hour and the 0.25 (dash), 0.5 (solid), 0.75 (dot) quantiles at each lag are plotted.}
    \label{fig:semi_var_west_2000}
\end{figure}

The semivariograms in Figure~\ref{fig:semi_var_west_2000} show that spatial dependence for the simulated data is lower than the observed NSRDB data at the same lags for the 0.25, 0.5, and 0.75 quantiles of the semivariances. The simulated data show similar agreement in trends but do not quite match the height of the semivariograms across all months analyzed. The smaller spatial autocorrelation at similar lags indicates there is a larger dependence at larger distances than what is seen in observed data. This lack of variability could be due to a variety of factors and is likely attributable to using four basis functions to incorporate cloud patterns and noise in simulated downscaled, even though we saw that four basis functions tends to capture a majority of the intra-daily variability across time. Here, the noise component results in smoother spatial patterns in the downscaled data compared to the observed data which may indicate the downscaled data does not spatially capture small scale cloud patterns, which are well documented by satellite data in the NSRDB. 

\subsection{Spatial downscaling}

\begin{figure}
    \centering
    \includegraphics[width=\textwidth]{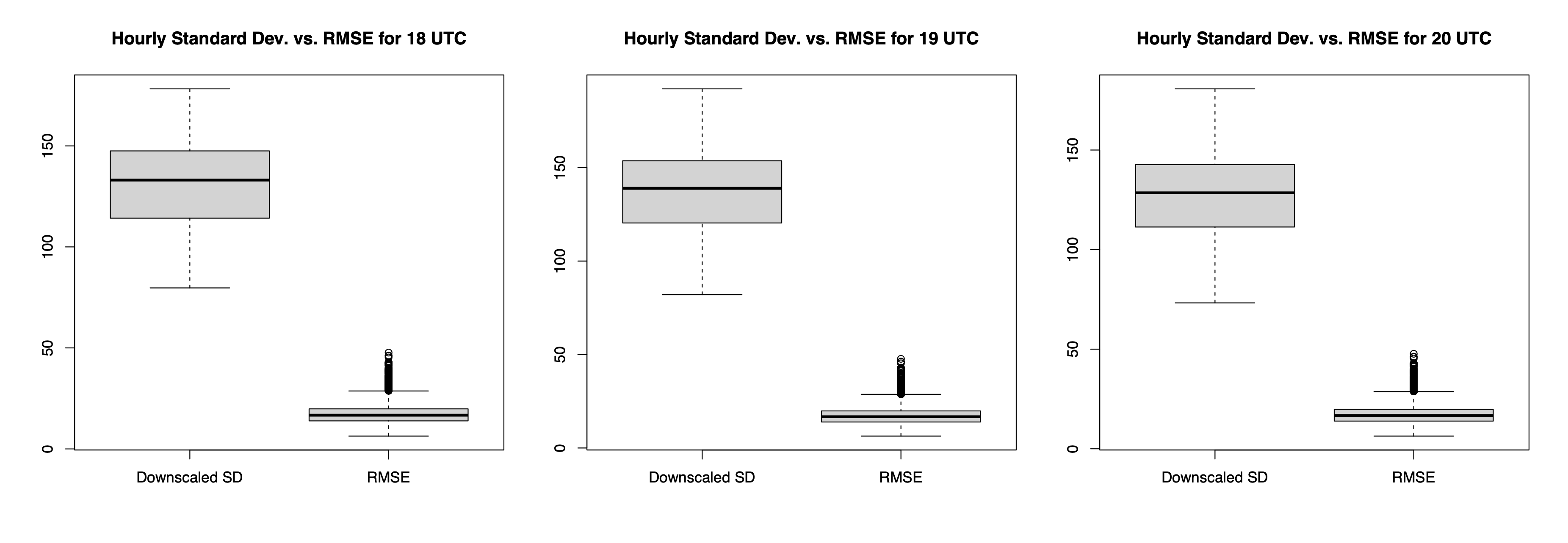}
    \caption{Standard deviation of GHI for three hours close to local solar noon in western CONUS compared to the RMSE between predicted and observed GHI for the same hours and region for days in January at the 8 km grid.}
    \label{fig:hrly_std_vs_rmse_sp}
\end{figure}

Variability in GHI greatly affects PV production, particularly for hours when there is a low solar zenith angle or hours close to solar noon. Therefore, the spatial downscaling method should maintain similar amounts of GHI variability to accurately model PV production at finer spatial scales. Comparing the hourly standard deviation and the RMSE indicates whether using TPS provides sufficient day to day variability, which will help with modeling PV production, or whether more complicated spatial downscaling methods may be necessary. Resulting RMSE values between downscaled data using TPS and observations are a fraction of the daily variability represented in a single hour across the month of January  for three hours surrounding local noon, or low zenith angle, as seen in Figure~\ref{fig:hrly_std_vs_rmse_sp}. This indicates that while TPS may not reproduce the exact structure of GHI at finer grids, the variability across days for high PV production hours of the downscaled data is much higher than the error for a set of locations. Therefore, downscaling to the 8 km grid using more complicated methods may not be necessary to understand day to day variability of GHI. One can rely on the distribution of hourly GHI simulations to get an accurate representation of day to day variability. Other months showed similar results for times when the sun is close to directly overhead in this region as well as central and eastern CONUS (see supplemental information). 

\section{Conclusions and future work}
\label{sec:conclusion}

This work introduces a new method to downscale daily average GHI values to hourly averages using functional principal components. We leverage the diurnal properties of solar radiation and propose splitting downscaling into two steps. First, we generate a diurnal template, estimated using clearsky GHI data, and adjust it according to given daily total GHI. We model the remaining diurnal cycle as a sum of basis functions whose coefficients have a distribution that depends on GHI and are spatially dependent. The model simulates fields of spatially correlated, hourly GHI data downscaled from daily average GHI and is efficient for moderately sized domains. We developed a local tile approach that allows for scaling up to continental scales and a variety of terrain. The method is novel in providing realistic spatially correlated fields and also reproducing the constraints inherent in solar radiation over a diurnal cycle. The resulting downscaled data closely aligns with summary statistics of observed data including daily total GHI, hourly time derivatives, proportions of clearsky GHI by hour, and cumulative distributions of GHI data by hour. Semivariograms for the downscaled data indicated higher dependence over larger distances, but generally followed similar trends as observed data.

Finally, we assess spatial downscaling using a simpler TPS method through comparing the hourly variability of GHI and RMSE for hours surrounding local solar noon. We found that day to day variability of downscaled data is higher than the RMSE between observed and predicted GHI on the 8 km grid. These results suggest that using TPS to spatially downscale GHI is sufficient for inference of day to day variability for times when the sun is higher in the sky and more complicated spatial downscaling methods may be unnecessary.

This work contributes to the production of a large data product containing hourly downscaled data for future years, up to 2100, at the 8 km level for CONUS. Future work includes applying these methods to RCMs for future years to produce hourly data with realistic simulations for solar radiation. These projections could then be used for site planning under various climate change scenarios. We anticipate this method to be available out-of-the-box for implementation with any RCM or, potentially, any GCM. Finally, this method was created specifically for solar radiation but could be adapted for other variables that have strong diurnal patterns and that are also spatially dependent, for example temperature. We hope that the method and its implementation on RCM data will provide end users with a clearer idea of the potential future for solar radiation to aid in renewable energy planning.

\bigskip

\newpage
\bibliographystyle{agsm}
\bibliography{building_high_res}

\end{document}